\documentclass[aps,pra,superscriptaddress, reprint, longbibliography]{revtex4-2}

\usepackage{xcolor}
\usepackage{graphicx}
\usepackage{dcolumn}
\usepackage{bm}
\usepackage{bbm}
\usepackage{relsize}
\usepackage{romannum}
\usepackage{csquotes}
\usepackage{xcolor}
\usepackage{soul}
\usepackage{floatrow}
\usepackage{amsmath}
\usepackage{tikz}
\usepackage[mathscr]{eucal}
\usepackage{physics}

\begin{document}
\preprint{APS/123-QED}

\title{Multistability and Self-Trapping in Cavity–Magnonic Dimer}
\author{Pooja Kumari Gupta}
 \affiliation{Department of Physics, Indian Institute of Technology Guwahati, Guwahati-781039, Assam, India}

\author{Amarendra K. Sarma}
\affiliation{Department of Physics, Indian Institute of Technology Guwahati, Guwahati-781039, Assam, India}
 
\author{Subhadeep Chakraborty}%
\email{c.subhadeep91@gmail.com}
\affiliation{Centre for Quantum Technologies, National University of Singapore, 3 Science Drive 2, Singapore 117543, Singapore}

\begin{abstract}
    We show that a driven–dissipative cavity–magnonic dimer supports multistability with coexisting symmetric and symmetry-broken steady states. The interplay between magnon Kerr nonlinearity and photon tunneling induces magnon self-trapping, leading to a persistent population imbalance between the two resonators. In the vicinity of saddle-node bifurcations, the system exhibits critical slowing down, with relaxation times far exceeding the intrinsic dissipation scale. Focusing on quantum correlations, we analyze the quantum fidelity and mutual information between the intercavity magnon modes. We find that both the infidelity and the mutual information increase sharply near the phase boundaries, providing clear quantum signatures of the multistable and symmetry-broken phases. Our results establish cavity magnonic dimers as a versatile platform for exploring nonlinear nonequilibrium physics in hybrid quantum systems.
\end{abstract}

\maketitle

\textit{Introduction:} Non-equilibrium phases in driven--dissipative quantum systems have attracted sustained interest owing to their ability to host collective phenomena far from equilibrium~\cite{polkovnikov2011colloquium, kessler2012dissipative, carusotto2013quantum, minganti2018spectral}.
Recent advances in hybrid quantum architectures~\cite{xiang2013hybrid,kurizki2015quantum} have enabled the controlled exploration of regimes where intrinsic nonlinear interaction competes with coherent driving and environmental dissipation, stabilizing non-equilibrium steady states that exhibit dissipative phase transitions~\cite{minganti2018spectral,beaulieu2025observation} and quantum critical behavior~\cite{angerer2017ultralong, brookes2021critical}. At the single-resonator level, this nonlinear response manifests as optical bistability, where a driven cavity supports multiple steady states distinguished by their excitation populations~\cite{drummond1980quantum,rempe1991optical}. Such physics has been widely explored across cavity and circuit QED platforms~\cite{carmichael2015breakdown, fitzpatrick2017observation}, Kerr resonators~\cite{casteels2016power, bartolo2016exact, casteels2017critical,rodriguez2017probing,chen2023quantum}, Rydberg gases~\cite{marcuzzi2014universal, de2016intrinsic}, and optomechanical systems~\cite{ghobadi2011quantum, bibak2023dissipative}.  
Coupling nonlinear resonators further introduces a spatial degree of freedom captured by the minimal Bose--Hubbard dimer, where the interplay between onsite interaction and tunneling can stabilize symmetry-broken steady states with uneven populations, giving rise to spontaneous symmetry breaking~\cite{cao2016two,casteels2017quantum,casteels2017optically, garbin2022spontaneous, xu2024phase} and self-trapping~\cite{albiez2005direct, levy2007ac, zibold2010classical, lagoudakis2010coherent, schmidt2010nonequilibrium, abbarchi2013macroscopic,vivek2023nonequilibrium, ray2024ergodic, vivek2025self}.

Here, we propose cavity magnonics~\cite{lachance2019hybrid, rameshti2022cavity, yuan2022quantum} as a versatile platform for investigating non-equilibrium phases in driven--dissipative settings. These systems rely on the coherent interaction between cavity photons and collective spin excitations (magnons) in ferromagnetic crystals. Their low damping rates, high tunability, and strong light--matter coupling have established cavity magnonics as a promising testbed for cavity--QED-like phenomena~\cite{goryachev2014high, zhang2015cavity, abdurakhimov2015normal, zare2015magnetic}. Moreover, magnons can hybridize with multiple degrees of freedom—including microwave~\cite{zhang2014strongly, PhysRevLett.113.083603, bai2015spin} and optical photons~\cite{osada2016cavity, zhu2020waveguide, bittencourt2022optomagnonics}, phonons~\cite{zhang2016cavity, potts2021dynamical, shen2022mechanical,xu2025kerr}, and superconducting qubits~\cite{tabuchi2015coherent, morris2017strong, wolski2020dissipation}—leading to a wide range of experimentally demonstrated effects such as magnon-induced transparency~\cite{wang2018magnon}, magnon dark modes~\cite{zhang2015magnon}, microwave–to-optical conversion~\cite{hisatomi2016bidirectional,zhu2020waveguide, ihn2020coherent}, non-hermitian exceptional points~\cite{harder2017topological, zhang2017observation}. Nonlinear cavity magnonics~\cite{zheng2023tutorial} further provides access to bistability~\cite{wang2018bistability, shen2022mechanical} and multistability~\cite{shen2021long} arising from the intrinsic Kerr nonlinearity, induced by magnetocrystalline anisotropy~\cite{gurevich1984magnetization, stancil2009spin}. Notably, such nonlinear interactions between magnons and complementary subsystems, such as cavity photons, phonons, and superconducting qubits, offer unprecedented controllability for engineering nonclassical magnonic states, with potential applications in quantum technologies. In this context, several theoretical works have proposed, such as nonclassical state generation~\cite{bittencourt2019magnon, sharma2021spin, yuan2020magnon}, magnon squeezing~\cite{li2019squeezed,yang2021bistability, PhysRevA.110.063504}, and hybrid entanglement~\cite{li2018magnon,li2019entangling, zhang2019quantum, yuan2020enhancement, yuan2020enhancement, yuan2020steady}.

Motivated by these advances, we investigate the emergence of non-equilibrium steady states and critical behavior in a driven dissipative cavity--magnon dimer. While coupled magnonic dimers have primarily been studied in symmetric configurations~\cite{hidki2022quantifying,peng2025symmetric}, we extend this framework to the asymmetric regime and demonstrate the appearance of multistability and self-trapped states characterized by a persistent magnon population imbalance. Notably, the required nonlinearities lie within current experimental reach~\cite{shen2022mechanical}. However, far fewer studies have addressed the behavior of quantum correlations in the vicinity of nonequilibrium critical points, particularly in multistable and symmetry-broken regimes. Very recently, driven-dissipative Dicke-type systems have shown that quantum correlations can exhibit pronounced signatures near phase transitions~\cite{soldati2021multipartite, boneberg2022quantum}, highlighting the importance of fluctuation effects beyond mean-field descriptions. Here, we analyze the quantum fidelity and mutual information between the intercavity magnon modes and identify sharp enhancements near the phase boundaries, revealing clear quantum signatures of the multistable and symmetry-broken transitions.

\begin{figure}[!t]
    \centering
    \includegraphics[width=0.98\textwidth]{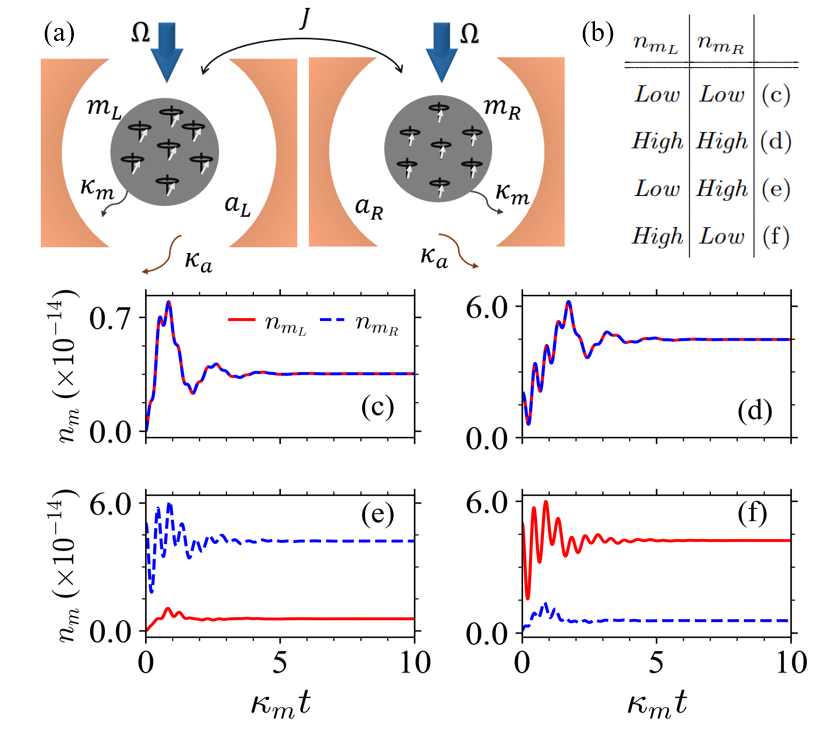}\\
     \caption{(a) Schematic diagram of the driven cavity--magnon dimer. (b) Steady-state magnon occupation for different dynamical classes. (c)--(f) Semiclassical dynamics of the cavity--magnon dimer reaching different steady states. Symmetric initial conditions evolve toward symmetric steady states with either both high or both low magnon populations, whereas asymmetric initial conditions converge to symmetry-broken steady states characterized by high--low or low--high populations in the respective modes. The parameters used in the simulation are $\omega_{a}/2\pi = 10$~GHz, $\kappa_{a}/2\pi = \kappa_{m}/2\pi = 1$~MHz, $\Delta_{a}/2\pi = \Delta_{m}/2\pi = -11$~MHz, $K/2\pi = 9$~nHz, $g/2\pi = 7$~MHz, $J=0.8\kappa_{a}$, $P_d=30$~mW.}
    \label{fig:model}
\end{figure}

\textit{Model and semiclassical analysis:} A schematic of the proposed cavity--magnonic dimer (CMD) is shown in Fig.~\ref{fig:model}(a). The system comprises two microwave cavities, each hosting a yttrium iron garnet (YIG) sphere and mutually coupled via photon tunneling at rate $J$. The intra-cavity photon fields are described by bosonic operators $a_i^\dagger$ and $a_i$ ($i=L,R$), while the corresponding magnon excitations are represented by $m_i^\dagger$ and $m_i$. The resonance frequencies of the cavity and magnon are denoted by $\omega_a$ and $\omega_m$, respectively. Photons couple to magnons via a magnetic dipole interaction of strength \(g\), and \(K\) denotes the magnon-Kerr nonlinearity.

To realize a driven--dissipative scenario, both magnon modes are coherently driven by identical microwave fields of power $P_d$ and frequency $\omega_d$, while photon and magnon losses occur at rates $\kappa_a$ and $\kappa_m$, respectively. In the rotating frame of the drive, the semiclassical dynamics read
\begin{equation}
\label{eqn:QLE}
\begin{aligned}
    \dot{a}_{i} &= -(\iota \Delta_{a} + \kappa_{a})a_{i} - \iota g m_{i} + \iota J a_{\bar{i}}, \\
    \dot{m}_{i} &= -(\iota \Delta_{m} + \kappa_{m})m_{i} - \iota g a_{i} - 2\iota K |m_{i}|^{2} m_{i} + \Omega ,
\end{aligned}
\end{equation}
where $\Delta_a=\omega_a-\omega_d$ and $\Delta_m=\omega_m-\omega_d$ denote the cavity and magnon detunings, and $\Omega=\sqrt{2\kappa_m P_d/\hbar\omega_d}$ is the drive amplitude. Notably, the dimer exhibits exchange symmetry: under the parity operation $\mathcal{O}_{L}\leftrightarrow\mathcal{O}_{R}$, the equations of motion remain invariant.

In the strong-driving limit, the operators are well described by complex fields, $a_i=\sqrt{n_{a_i}}e^{\iota\psi_i}$ and $m_i=\sqrt{n_{m_i}}e^{\iota\phi_i}$, where $n$ represents the amplitudes and $\psi,\phi$ denotes the corresponding phases, respectively. The resulting amplitude and phase dynamics are
\begin{equation}
\label{eqn:n_theta}
\begin{aligned}
 \dot{n}_{a_i} &= -\kappa_{a} n_{a_i} 
                 - g \sqrt{n_{a_i} n_{m_i}} \sin(\psi_i- \phi_i ) \\
               &\quad - J\sqrt{n_{a_i} n_{a_{\bar{i}}}} \sin(\psi_{\bar{i}} - \psi_i), \\
 \dot{n}_{m_i} &= -\kappa_{m} n_{m_i} 
                 + g \sqrt{n_{a_i} n_{m_i}} \sin (\psi_i - \phi_i)
                 + \Omega \sqrt{n_{m_i}} \cos\phi_i, \\
 \dot{\psi}_i &= -\Delta_a 
                - g \sqrt{\tfrac{n_{m_i}}{n_{a_i}}} \cos (\psi_i-\phi_i )
                + J \sqrt{\tfrac{n_{a_{\bar{i}}}}{n_{a_i}}} \cos (\psi_{\bar{i}} - \psi_i), \\
 \dot{\phi}_i &= -\Delta_{m_i}^{\prime} 
                - g \sqrt{\tfrac{n_{a_i}}{n_{m_i}}} \cos (\psi_i - \phi_i)
                - \tfrac{\Omega}{\sqrt{n_{m_i}}}\sin\phi_i.
\end{aligned}
\end{equation}
Here, $\Delta^{\prime}_{m_i} = \Delta_{m_i}+2K\vert\langle m_i\rangle\vert^2$ denotes the effective magnon detuning.

To characterize the dynamics, we look for the steady-state solutions and analyze their stability. Solving Eq.~\eqref{eqn:n_theta} yields coupled cubic equations for the magnon populations,
\begin{equation}
\label{eqn:multistability}
\begin{aligned}
    4K^2 n_{m_L}^3 + 4K\tilde{\Delta}_{m_L} n_{m_L}^2 
    + (\tilde{\Delta}_{m_L}^2 + \tilde{\kappa}_{m_L}^2)n_{m_L}
    - \Omega^2 = 0, \\
    4K^2 n_{m_R}^3 + 4K\tilde{\Delta}_{m_R} n_{m_R}^2 
    + (\tilde{\Delta}_{m_R}^2 + \tilde{\kappa}_{m_R}^2)n_{m_R}
    - \Omega^2 = 0,
\end{aligned}
\end{equation}
where $\tilde{\Delta}_{m_i}=\Delta_m-\eta_i\tilde{\Delta}_i$,  
$\tilde{\kappa}_{m_i}=\kappa_m-\sqrt{\eta_i g^2-\eta_i^2\tilde{\Delta}_i^2}$,  
$\eta_i=g^2/(\kappa_i^2+\tilde{\Delta}_i^2)$,  
$\tilde{\Delta}_L=\Delta_a-Jf\cos(\Delta\psi)$,  
$\tilde{\Delta}_R=\Delta_a-J\cos(\Delta\psi)/f$,  
$\kappa_L=\kappa_a+Jf\sin(\Delta\psi)$,  
$\kappa_R=\kappa_a-J\sin(\Delta\psi)/f$,  
$\Delta\psi=\psi_R-\psi_L$, and $f=n_{a_R}/n_{a_L}$.

Next, we classify the steady states into two classes: (i) \textit{symmetric}, defined by 
$n_{a_L/m_L}=n_{a_R/m_R}$, $\psi_L=\psi_R$, and $\phi_L=\phi_R$; and (ii) \textit{asymmetric}, 
characterized by unequal amplitudes and phases. Homogeneous driving eliminates the 
antisymmetric mode ($\Delta\psi=\pi$). Restricting the dynamics to the symmetric subspace 
($\Delta\psi=0$, $f=1$) reduces the system to an effective single cavity--magnon bistability equation~\cite{yang2021bistability, PhysRevA.110.063504},
\begin{equation}
\label{eqn:bistability}
4 K^{2} n_{m}^{3} + 4 \Delta_{0} K n_{m}^{2} + (\Delta_{0}^{2} + \kappa_{0}^{2}) n_{m} - \Omega^{2} = 0,
\end{equation}
with the bistability criterion $\Delta_{0}^{2}-3\kappa_{0}^{2}>0$. Here, the effective single cavity parameters are defined as follows:  
$\Delta_{0}=\Delta_m-\eta(\Delta_a-J)$,  
$\kappa_{0}=\kappa_m-\eta\kappa_a$, and  
$\eta=g^2/(\kappa_a^2+(\Delta_a-J)^2)$.  
Note that the effective detuning, \( (\Delta_a - J) \), is determined solely by the difference between the cavity detuning and the hopping strength. Consequently, the system exhibits identical behavior for different hopping strengths, provided that the difference remains fixed. 

\begin{figure}[!t]
    \centering
    \textbf{(a)}\\
    \includegraphics[width=0.8\textwidth]{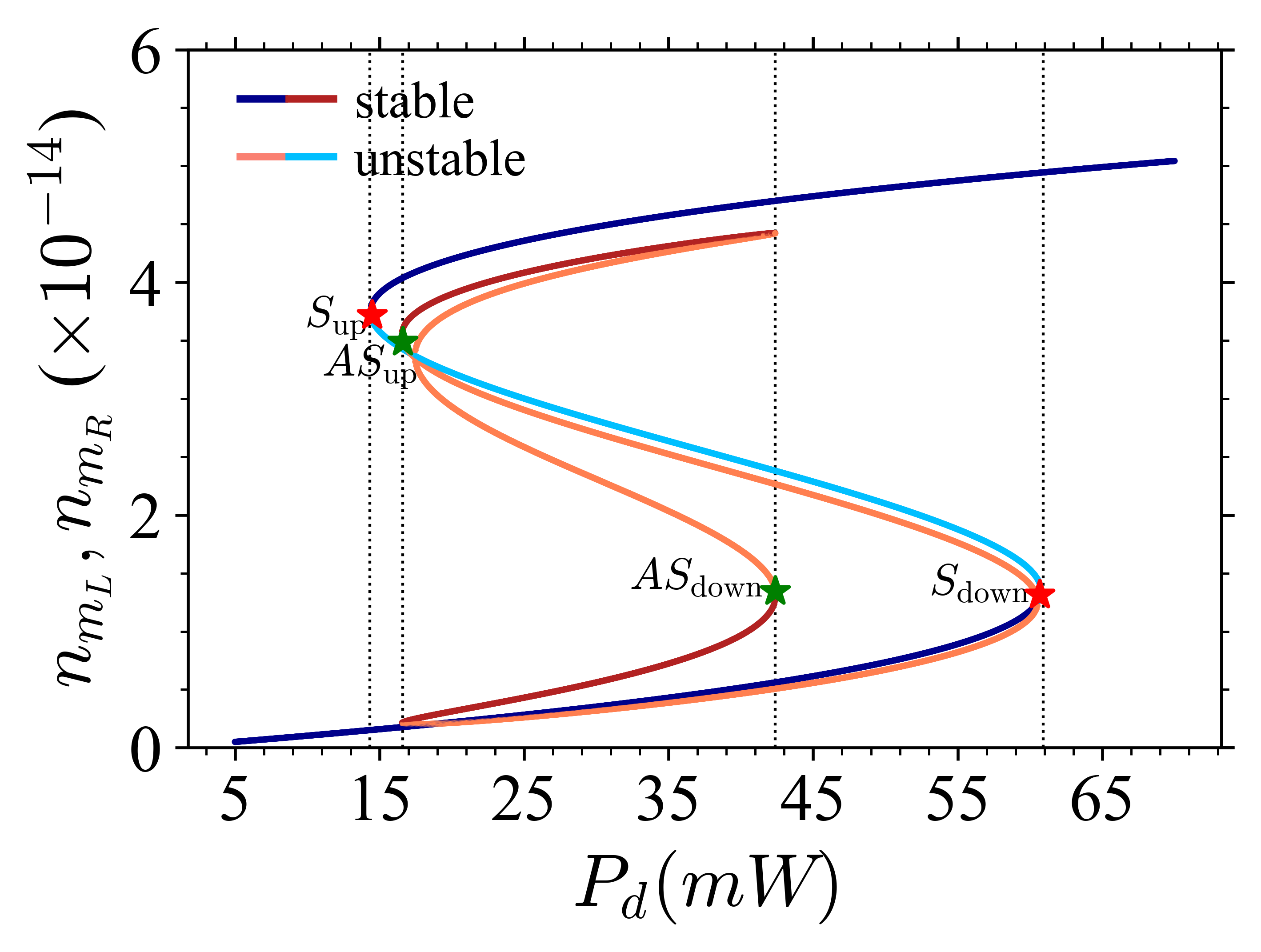}
    
    \vspace{0.1cm}
    \textbf{(b)}\\
    \includegraphics[width=0.87\textwidth]{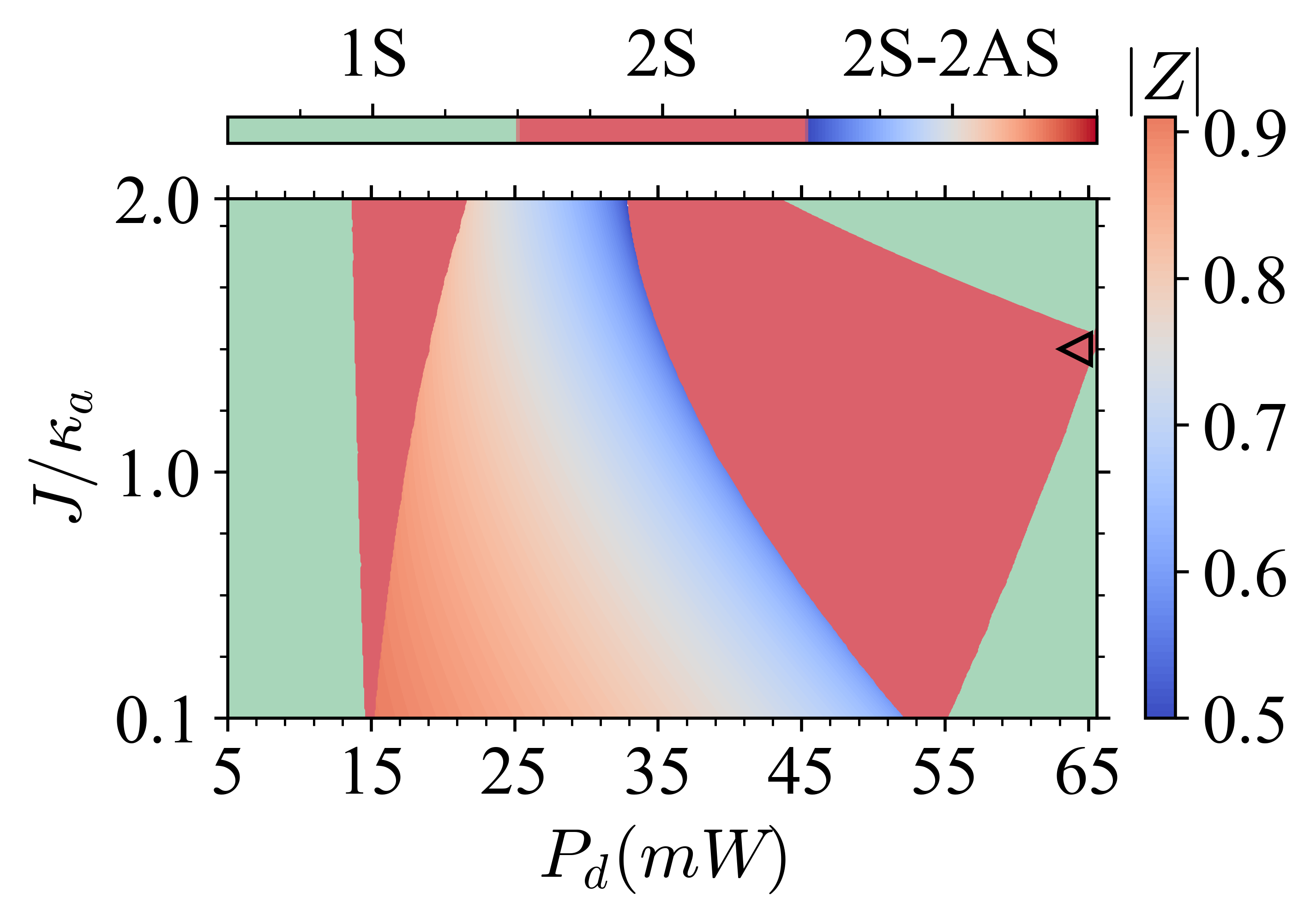}
    
    \caption{(a) Steady-state magnon numbers $n_{m_L}$ and $n_{m_R}$ versus drive power $P_d$(mW). 
    Red markers denote symmetric saddle-node bifurcation points, and green markers denote asymmetric ones. 
    (b) Phase diagram of the symmetry-broken steady states in the $P_d$–$J/\kappa_a$ plane. 
    1S: monostable symmetric phase; 2S: bistable symmetric phase; 2S–2AS: multistable phase supporting four coexisting steady states (two symmetric and two asymmetric). 
    The corresponding steady-state population imbalance $\vert Z\vert$ is also shown. The $\triangleleft$ symbol marks the emergence of the Hopf-bifurcation phase.}
    \label{fig:stability}  
\end{figure}

Fig.~\ref{fig:stability}(a) shows the steady state magnon population $n_m$ versus drive power $P_d$. The symmetric branch exhibits the characteristic S-shaped bistability, with two stable outer branches separated by an unstable one. In contrast, the asymmetric solutions split into two S-shaped curves, each undergoing a saddle-node (SN) bifurcation at either its lower or upper branch. 
Together, they yield one low- and one high-amplitude asymmetric steady state. 
Consequently, over the interval $AS_{\mathrm{up}} \leq P_d \leq AS_{\mathrm{down}}$, the dimer supports four coexisting attractors: two symmetric (\textit{low--low}, \textit{high--high}) and two asymmetric (\textit{low--high}, \textit{high--low}) steady states (see Fig.~\ref{fig:model}(b)).
This marks the multistablity phase, arising from the coupling of two individually bistable cavity--magnon subsystems.
Therefore, the long-time dynamics is selected by the basin of attraction, despite homogeneous driving (see Fig.~\ref{fig:model}(c)--(f)).

Fig.~\ref{fig:stability}(b) presents the complete phase diagram of the dimer in the tunneling–drive plane. 
The phase boundaries delineate regions of mono-, bi-, and multi-stable phases, marking the parameter ranges that support one (1S), two (2S), and four (2S-2AS) stable steady states, respectively. 
The bistable boundaries exhibit an approximately linear dependence on the tunneling strength. 
Interestingly, the symmetry-broken regime always appears well inside this bistable domain. As $J$ increases, the asymmetric sector progressively narrows and straightens upward, as stronger photon tunneling favors population equilibration between the cavities~\cite{cao2016two}. A defining hallmark of these asymmetric steady states is magnon self-trapping, reminiscent of macroscopic self-trapping in Josephson junction platforms~\cite{albiez2005direct, levy2007ac, zibold2010classical, lagoudakis2010coherent, schmidt2010nonequilibrium, abbarchi2013macroscopic, vivek2023nonequilibrium, ray2024ergodic, vivek2025self}.  
As a metric, we quantify this effect via the steady-state population imbalance $|Z| = |(n_{m_L} - n_{m_R})/(n_{m_L} + n_{m_R})|$. We observe a finite steady-state imbalance between the magnon modes even under homogeneous driving and dissipation. 
The imbalance is largest near the upper SN bifurcation points of the asymmetric branch, where the difference between high- and low-population states is most pronounced. 
It then decreases monotonically toward the lower SN point, indicating gradual population equilibration, and ultimately vanishes once the system enters the symmetric bistable phase. In passing, we note that in Fig.~\ref{fig:stability}(b) the $\triangleleft$ symbols mark the emergence of a Hopf-bifurcation phase, which we do not discuss here and leave for future work. 

\begin{figure}[!t]
    \centering
    \includegraphics[width=\textwidth]{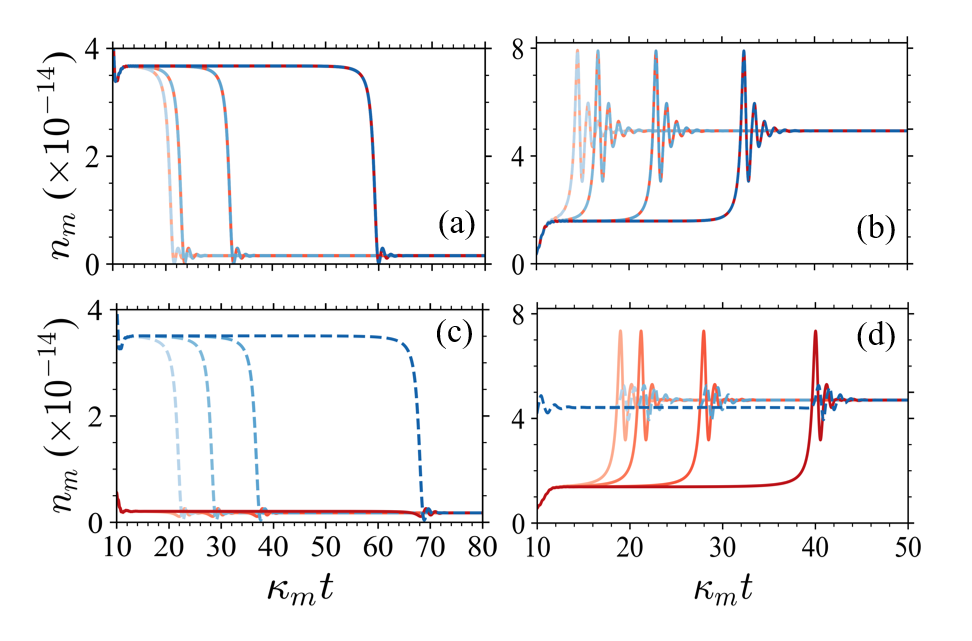}
    \caption{Quench dynamics of the magnon numbers $n_{m_L}$ (solid red) and $n_{m_R}$ (dashed blue) for different drive powers near the saddle-node bifurcation points marked in Fig.~\ref{fig:stability}: (a) $S_{\mathrm{up}}$, (b) $S_{\mathrm{down}}$, (c) $AS_{\mathrm{up}}$, and (d) $AS_{\mathrm{down}}$. 
    In all panels, the transient time increases from the lighter to darker curves, with the darkest curve corresponding to the point closest to the bifurcation.}
    \label{fig:CSD}
    \end{figure}

To investigate the dynamical response of the cavity--magnon dimer near criticality, we analyze the phenomenon of critical slowing down (CSD)~\cite{hohenberg1977theory, bonifacio1979critical, angerer2017ultralong, krimer2019critical, beaulieu2025observation,brookes2021critical,krimer2019critical} using quench dynamics in the vicinity of the saddle-node bifurcation points identified in Fig.~\ref{fig:stability}(a). Fig.~\ref{fig:CSD}(a,b) correspond to the symmetric bifurcation points $S_{\mathrm{up}}$ and $S_{\mathrm{down}}$, while Figs.~\ref{fig:CSD}(c,d) depict the dynamics near their asymmetric counterparts $AS_{\mathrm{up}}$ and $AS_{\mathrm{down}}$.
For the upper symmetric bifurcation $S_{\mathrm{up}}$, the system is first prepared in a steady state by evolving it at a drive power well above the critical value. The drive is then abruptly quenched to a value slightly below the threshold, after which the system relaxes toward the attractor associated with the lower symmetric branch. An analogous protocol is employed near $S_{\mathrm{down}}$: the system is initialized far below the critical power and subsequently quenched to a value just above it, leading to relaxation onto the upper symmetric branch. 
The same strategy is applied to the asymmetric bifurcation points $AS_{\mathrm{up}}$ and $AS_{\mathrm{down}}$. Notably, after the quench, the dynamics invariably converge to symmetric attractors. This occurs because the quench places the system in a parameter regime where the asymmetric solutions cease to exist. Since asymmetric states emerge in pairs, the termination of one branch at the saddle-node simultaneously eliminates its counterpart, leaving the symmetric state as the only accessible attractor.
Across all cases, the system exhibits a pronounced transient bottleneck before reaching the final steady state. As the quench amplitude approaches the critical value, the relaxation time increases sharply, consistent with a power-law divergence characteristic of CSD. The resulting evolution unfolds over timescales far exceeding the intrinsic dissipation scale \(1/\kappa\) ($\kappa=\kappa_a =\kappa_m$).

\textit{Quantum fluctuations:} Although the semiclassical analysis successfully captures multistability and critical slowing down, it is equally important to examine the influence of quantum fluctuations near the phase boundaries. To this end, we analyze Gaussian fluctuations around the stable semiclassical fixed points. Let the fixed-point vector be 
$\vec{\chi}^* = (X^*_{a_L}, Y^*_{a_L}, X^*_{m_L}, Y^*_{m_L}, X^*_{a_R}, Y^*_{a_R}, X^*_{m_R}, Y^*_{m_R})^T$, 
and define the vector of associated fluctuation operators as 
$\hat{u}(t) = (\delta \hat{X}_{a_L}, \delta \hat{Y}_{a_L}, \delta \hat{X}_{m_L}, \delta \hat{Y}_{m_L}, \delta \hat{X}_{a_R}, \delta \hat{Y}_{a_R}, \delta \hat{X}_{m_R}, \delta \hat{Y}_{m_R})^T$. 
Each cavity and magnon operator is expressed in the quadrature basis according to
$\delta X_{o_i} = (\delta \hat{\mathcal{O}} + \delta \hat{\mathcal{O}}^{\dagger})/\sqrt{2}$ and 
$\delta Y_{o_i} = (\delta \hat{\mathcal{O}} - \delta \hat{\mathcal{O}}^{\dagger})/(\iota\sqrt{2})$. 
The system operators are then decomposed as 
$\hat{\mathcal{O}} = \vec{\chi}^* + \hat{u}$, 
which is valid in the strong-drive regime where the fluctuations remain small, i.e., $\|\hat{u}\| \ll 1$.

Linearizing Eq.~\eqref{eqn:QLE} about $\vec{\chi}^*$ yields the quantum Langevin equation
$\dot{\hat{u}}(t) = A \hat{u}(t) + \hat{\xi}(t)$, 
where 
$
\hat{\xi}(t) =
\left[
\sqrt{2\kappa_a} X^{\mathrm{in}}_{a_L}, 
\sqrt{2\kappa_a} Y^{\mathrm{in}}_{a_L}, 
\sqrt{2\kappa_m} X^{\mathrm{in}}_{m_L}, 
\sqrt{2\kappa_m} Y^{\mathrm{in}}_{m_L}, 
\sqrt{2\kappa_a} X^{\mathrm{in}}_{a_R}, \right.\\ \left.
\sqrt{2\kappa_a} Y^{\mathrm{in}}_{a_R}, 
\sqrt{2\kappa_m} X^{\mathrm{in}}_{m_R}, 
\sqrt{2\kappa_m} Y^{\mathrm{in}}_{m_R} 
\right]^T
$ denotes the vector of input noises. The corresponding input noise operators have zero mean and satisfy the correlations 
$\langle a^{\mathrm{in}}(t)a^{\mathrm{in},\dagger}(t')\rangle = \delta(t-t')$ and 
$\langle m^{\mathrm{in}}(t)m^{\mathrm{in},\dagger}(t')\rangle = \delta(t-t')$.
Here, the drift matrix assumes the block form
\begin{equation}
A = I_2 \otimes A_{am} 
+ \sigma_x \otimes 
\left[
\begin{pmatrix}
0 & -J\\
J & 0
\end{pmatrix}
\oplus 0_2
\right],
\end{equation}
indicating that photon tunneling couples the two cavities exclusively within the photonic quadrature subspace, while the magnon sector remains uncoupled, as represented by the null block. 
The single cavity--magnon block is given by
\begin{equation}
A_{a_i m_i}=
\begin{pmatrix}
-\kappa_{a} & \Delta_{a} & 0 & g\\
-\Delta_{a} & -\kappa_{a} & -g & 0\\
0 & g & -\kappa_{m}+\Delta^y_{K_i} & \Delta^{\prime\prime}_{m_i}-\Delta^x_{K_i}\\
-g & 0 & -\Delta^{\prime\prime}_{m_i}-\Delta^x_{K_i} & -\kappa_{m}-\Delta^y_{K_i}
\end{pmatrix},
\end{equation}
with $\Delta^{\prime\prime}_{m_i}=\Delta_{m_i}^\prime+2K|\langle m_i\rangle|^2$ and $\Delta_{K_i}=2K\langle m_i\rangle^2=\Delta^x_{K_i}+i\Delta^y_{K_i}$.

Due to the linearized dynamics and the zero-mean Gaussian nature of the quantum noises, the cavity--magnon dimer is fully characterized by an $8\times8$ covariance matrix (CM) with elements $V_{ij}=\langle \hat{u}_i(t)\hat{u}_j(t)+\hat{u}_j(t)\hat{u}_i(t)\rangle/2$. The steady-state CM follows from the Lyapunov equation~\cite{sarma2021continuous}
\begin{equation}
\label{eqn:cov_dynamics}
AV+VA^{T}=-D,
\end{equation}
where $D=\mathrm{diag}(\kappa_{a},\kappa_{a},\kappa_{m},\kappa_{m},\kappa_{a},\kappa_{a},\kappa_{m},\kappa_{m})$ is the diffusion matrix. To quantify correlations between the two subsystems, we extract the reduced covariance matrix (CM) of the left and right magnon modes. The resulting bipartite Gaussian state is conveniently written in block form as $V^{(2)} = (\alpha \;\beta;\, \beta^{T}\; \gamma)$~\cite{olivares2012quantum, adesso2014continuous,sarma2021continuous}, where the $2\times2$ matrices $\alpha$ and $\gamma$ describe the local magnon fluctuations, while $\beta$ encodes the inter-magnon correlations.

\begin{figure}[!t]
    \centering
    \includegraphics[width=0.8\textwidth]{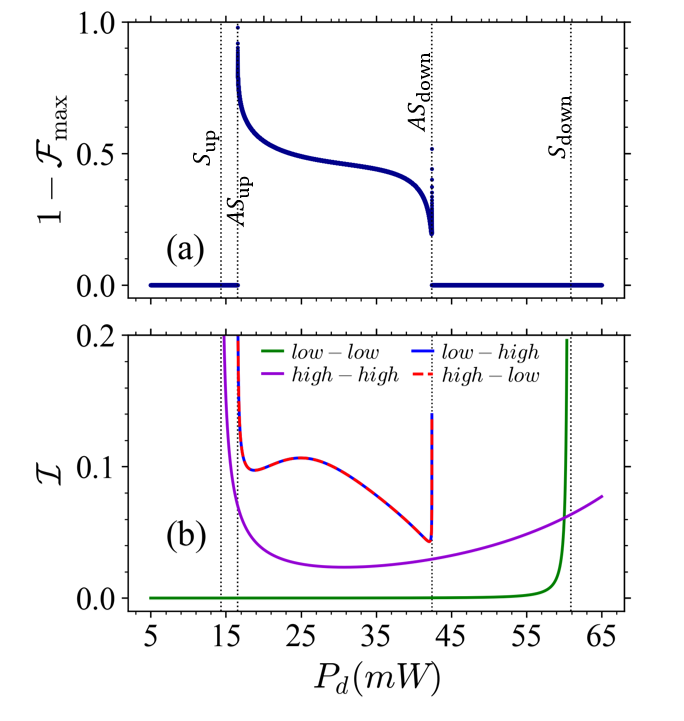}
    \caption{Variation of (a) maximum infidelity and (b) mutual information between the magnon modes in the left and right cavities as a function of the drive power $P_d$ (mW). 
    Panel (b) includes all combinations of low- and high-occupation states.}
    \label{fig:fidelity}
    \end{figure}

To quantify the overlap between the left and right magnon modes, we evaluate the quantum fidelity, maximized over accessible steady states~\cite{minganti2018spectral}. For two single-mode Gaussian states with covariance matrices $\alpha$ and $\gamma$ and vanishing first moments, the fidelity~\cite{banchi2015quantum} is given by $
\mathcal{F}(\alpha, \gamma)= \left(\sqrt{\delta + \Lambda} - \sqrt{\Lambda}\right)^{-1}$, where $\delta = \det(\alpha + \gamma)$ and $\Lambda = 4\, \det\!\left(\alpha + \frac{\iota}{2} \mathcal{T}\right)\det\!\left(\gamma + \frac{\iota}{2} \mathcal{T}\right)$. Here, $\mathcal{T}$ is defined as $\mathcal{T}_{lm} = - \iota [ \hat{u}_l, \hat{u}_m ]$. This fidelity is symmetric, $\mathcal{F}(\alpha,\gamma)=\mathcal{F}(\gamma,\alpha)$, bounded between 0 and 1, and equals unity if and only if $\alpha=\gamma$.

Fig.~\ref{fig:fidelity}(a) shows the infidelity $(1-\mathcal{F})$ between the magnon modes in different stability regions as a function of drive power $P_d$. In the symmetric regime, the infidelity vanishes, indicating complete overlap between the intercavity magnon modes even in the presence of quantum noise. Within the multistable region, symmetry-broken states emerge, resulting in finite infidelity near the asymmetric saddle-node bifurcation points~\cite{casteels2017optically, minganti2018spectral}. While this metric clearly distinguishes symmetric from asymmetric phases, it does not resolve the upper and lower symmetric branches. To address this limitation, we analyze the quantum mutual information across the dimer. Notably, although each cavity--magnon subsystem is locally entangled due to the intrinsic Kerr nonlinearity~\cite{yang2021bistability, PhysRevA.110.063504}, we find no entanglement between the intercavity magnon modes. 

In what follows, we compute the quantum mutual information~\cite{olivares2012quantum} between the two magnon modes, defined as $\mathcal{I}(\alpha\!:\!{\gamma}) = f(\nu_\alpha) + f(\nu_\gamma) - f(\nu_+) - f(\nu_-)$, where $\nu_\alpha$ and $\nu_\gamma$ are the symplectic eigenvalues of the single-mode covariance matrices $\alpha$ and $\gamma$, while $\nu_{\pm} = \sqrt{(\Delta \pm \sqrt{\Delta^2 - 4 \det V^{(2)}})/2}$ with $\Delta = \det \alpha + \det \beta + 2\det \gamma$. The entropy function is $f(x) = \left(x+\tfrac{1}{2}\right)\ln\!\left(x+\tfrac{1}{2}\right) - \left(x-\tfrac{1}{2}\right)\ln\!\left(x-
\tfrac{1}{2}\right)$.
Fig.~\ref{fig:fidelity}(b) shows $\mathcal{I}$ as a function of drive power $P_d$ for all the accessible steady states. The mutual information respects the exchange symmetry of the dimer and also able to differentiates the symmetric \textit{low--low} and \textit{high--high} branches from the asymmetric configurations. Within the multistable region, asymmetric states exhibit larger correlations than the symmetric high-population branch, whereas the \textit{low--low} state remains nearly uncorrelated due to the effective weak nonlinear interaction. Both infidelity and mutual information grow sharply near the phase boundaries~\cite{casteels2017quantum}, reflecting enhanced fluctuations as the eigenvalues of the drift matrix approach zero. Such behavior is consistent with the emergence of a first-order dissipative phase transition associated with bistability.

\textit{Conclusion:} In conclusion, we investigated a driven--dissipative cavity--magnon dimer and uncovered a rich landscape of non-equilibrium steady states. Despite homogeneous driving, the interplay of Kerr nonlinearity and photon tunneling generates multistability with coexisting symmetric and symmetry-broken attractors, leading to magnon self-trapping. Near the saddle-node bifurcations, the system exhibits pronounced critical slowing down with relaxation times exceeding the intrinsic dissipation scale. Quantum correlations, quantified through fidelity and mutual information, further reveal enhanced fluctuations and intrercavity mode correlation close to the transition points. These results establish the cavity--magnon dimer as a versatile platform for exploring nonlinear nonequilibrium phenomena in hybrid quantum systems and provide a stepping stone toward larger magnonic arrays.

\textit{Acknowledgment:}
P.K.G. gratefully acknowledges a research fellowship from MoE, Government of India. A.K.S.  acknowledges the grant from MoE, Government of India (Grant No. MoE-STARS/STARS- 2/2023-0161).
\bibliographystyle{apsrev4-2}
\bibliography{CMD_references}
\end{document}